\documentclass[twocolumn]{aastex63}
\usepackage[encapsulated]{CJK}
\usepackage{mathrsfs}
\usepackage{subfigure}
\usepackage{epstopdf}
\usepackage{amsmath}
\usepackage{longtable}
\usepackage{booktabs}
\usepackage{hyperref} %\usepackage{threeparttable}
\usepackage{overpic}
\def\etal {et al.~}

\newbox\grsign \setbox\grsign=\hbox{$>$} \newdimen\grdimen \grdimen=\ht\grsign
\newbox\laxbox \newbox\gaxbox
\setbox\gaxbox=\hbox{\raise.5ex\hbox{$>$}\llap
     {\lower.5ex\hbox{$\sim$}}}\ht1=\grdimen\dp1=0pt
\setbox\laxbox=\hbox{\raise.5ex\hbox{$<$}\llap
     {\lower.5ex\hbox{$\sim$}}}\ht2=\grdimen\dp2=0pt
\shorttitle{Local spiral structure traced by Red Clump stars}
\shortauthors{Lin \etal}

\definecolor{malachite}{rgb}{0.34, 0.7, 0.22}

\begin{document}

   \title{Local spiral structure traced by Red Clump stars}
   
   \correspondingauthor{Ye Xu}
   \email{xuye@pmo.ac.cn}
   
   \author{Zehao lin}
   \author{Ye Xu}
   \affiliation{Purple Mountain Observatory, Chinese Academy of Sciences, Nanjing 210008, People's Republic of China}
   \affiliation{University of Science and Technology of China, 96 Jinzhai Road, Hefei 230026, People's Republic of China}
   \author{Ligang Hou}
   \affiliation{National Astronomical Observatories, Chinese Academy of Sciences, 20A Datun Road, Chaoyang District Beijing 100101, PR China}
   \author{Dejian Liu}
   \affiliation{Purple Mountain Observatory, Chinese Academy of Sciences, Nanjing 210008, People's Republic of China}
   \affiliation{University of Science and Technology of China, 96 Jinzhai Road, Hefei 230026, People's Republic of China} 
   \author{Yingjie Li}
   \affiliation{Purple Mountain Observatory, Chinese Academy of Sciences, Nanjing 210008, People's Republic of China}
   \author{Chaojie Hao}  
   \affiliation{Purple Mountain Observatory, Chinese Academy of Sciences, Nanjing 210008, People's Republic of China}
   \affiliation{University of Science and Technology of China, 96 Jinzhai Road, Hefei 230026, People's Republic of China} 
   \author{Jingjing Li}
   \affiliation{Purple Mountain Observatory, Chinese Academy of Sciences, Nanjing 210008, People's Republic of China}
   \author{Shuaibo Bian}  
   \affiliation{Purple Mountain Observatory, Chinese Academy of Sciences, Nanjing 210008, People's Republic of China}
   \affiliation{University of Science and Technology of China, 96 Jinzhai Road, Hefei 230026, People's Republic of China} 

\begin{abstract}

Using the cross-matched data of {\it Gaia} EDR3 and the 2MASS Point Source Catalog,
a sample of RC stars with parallax accuracies better than 20\% is
identified and used to reveal the nearby spiral
pattern traced by old stars.
As shown in the overdensity distribution of RC stars, there is an arc-like feature extended from $l\sim$~90$^\circ$ to $\sim$~243$^\circ$, which passed close to the Sun. This feature is probably an arm segment traced by old stars, indicating the Galaxy potential in the vicinity of the Sun. By comparing to the spiral arms depicted by young objects, we found that there are considerable offsets between the two different components of Galactic spiral arms. The spiral arm traced by RC stars tends to have a larger pitch angle, hence a more loose wound pattern.
\end{abstract}

\keywords{Galaxy structure(622); Galaxy stellar content(621); Solar neighbourhood(1509)}

%\maketitle
%
%-------------------------------------------------------------------

\section{Introduction}
\label{intro}

As observers, we are deeply embedded in the Galactic disc, making it a
difficult issue in astronomy to accurately reveal the spiral pattern
of the Milky Way and understand its formation and evolution.
In comparison to the Milky Way Galaxy, external spiral galaxies which are face on to can be intuitively mapped, studies on which have promoted our
understanding of the Galactic spiral structure.
It is suggested that there may be two different components of spiral
arms for a spiral galaxy~\citep[][and references therein]{Dobbs_Baba_2014}, one component consists of the star formation
arms (SF-arms) traced by diffuse or dense interstellar gas, and also
young objects (e.g., massive OB stars, H{\sc ii} regions, young open
clusters), the other is composed of a spiral pattern indicated by the
potential of old stars (hereafter P-arms throughout this work).
Early views suggested that the optical (indicating the SF-arms) and
near-infrared (hereafter NIR, indicating the P-arms) morphologies of
spiral galaxies were decoupled~\citep[e.g.][]{Block_Puerari_1999}.
Later on, the Ohio State University (OSU) Bright Spiral Galaxy Survey
found a good correlation between the optical and NIR band morphologies
for most of their observed spiral galaxies, however, dramatic
differences were also noticed in some cases~\citep{Eskridge+2002}.
A simple picture may not be applied to all different types of spiral galaxies. 
With some understanding of the spiral galaxies, we look back on our own Milky Way. Is there any similarities and/or differences between two different spiral-arm components of the Milky Way through different tracers?
%What about the Milky Way?

Since the 1950s, much efforts have been devoted to depict the spiral
structure of the Milky Way through the distributions or kinematics of
OB stars~\citep{Morgan+1952,Morgan+1953}, neutral hydrogen
gas~\citep{Christiansen_Hindman1952,van+1954,Oort+1958,Bok_1959,Levine+2006}, and some other kinds of spiral tracers (e.g., H{\sc ii} regions, CO gas,
young open clusters and Cepheid variable stars~\citet{Skowron+2019,Poggio+2021}), most of which were
based on kinematic methods.
Until the Bar and Spiral Structure Legacy (BeSSeL) Survey
\citep[][]{Reid+2019}, which has obtained precise trigonometric
parallaxes~\citep[][]{Xu+2006} for about 200 high-mass star-forming
regions (HMSFRs), the four segments of SF-arms together with the Local
Arm and several arm spurs have been depicted. This picture of SF-arms
based on HMSFRs was also confirmed or extended by using other young
objects with precise parallax measurements, e.g. massive OB
stars~\citep{Xu+2018b,Xu+2021} and young open clusters~\citep[$<$
  20~Myr,][]{Hao+2021}.

In the meantime, a different morphology for the P-arms of the Milky
Way was proposed. The statistics of NIR and far-infrared (FIR)
emissions of the Galactic disc indicate a two-armed spiral pattern,
i.e. the Scutum-Centaurus Arm and the Perseus
Arm~\citep{Drimmel_2000,Drimmel_Spergel_2001}.
The other two arms (Sagittarius-Carina and Norma) could perhaps be
presented only in gas and young objects, as they were traced only by
the FIR emission.
Then, the Spitzer/\textit{GLIMPSE} mid-infrared survey also supports
that the Milky Way is a barred, two-armed spiral
galaxy~\citep{Benjamin+2003,Benjamin+2005,Churchwell+2009}, because a
large number of red clump stars detected by \textit{GLIMPSE} present
obvious convergence near the tangential directions of the
Scutum-Centaurus Arm, in contrast, the tangential directions
corresponding to the Sagittarius-Carina Arm and Norma Arm are
indecipherable.
Since the past few years, the morphological and kinematic properties
of spiral arms traced by stars in the vicinity of the Sun can be
explored with the aid of the ESA's {\textit Gaia} mission.
By using the {\it Gaia} second data release (DR2),
\citet{Miyachi+2019} studied the surface density distribution of a
number of turn-off stars. These stars with an age of $\sim$1~Gyr
exhibit a marginal arm-like overdensity in the Galactic longitude of
90$^\circ~<~l~<$ 190$^\circ$, which is close to the Local Arm traced
by gas and young objects, but with a slightly larger pitch angle.
\cite{Khoperskov+2020} identified six stellar density structures in a
guiding coordinate space based on the {\it Gaia} DR2 sources having 6D
phase-space coordinates. These structures were interpreted as features
related to spiral arms and resonances.
Meanwhile, \cite{Hunt+2020} presented a different viewpoint. They
suggested that the stellar density structures found by
\cite{Khoperskov+2020} are known kinematic moving groups, rather than
spiral arms.
\citet{Poggio+2021} derived the density distribution of a large number
of upper main sequence stars~\citep[O, B, and A-type,][]{Poggio+2018},
which cover a wide age range. The resulting maps exhibit extended
arm-like structures indicating the arm segments in the Solar
neighborhood.
In addition, \citet{Hao+2021} confirmed that the distribution of old
open clusters (ages $>$ 1~Gyr) in the vicinity of the Sun does not
present any obvious arm-like features.

The difference between the SF-arms and P-arms is a fundamental
characteristic of the Milky Way, closely related to the formation
mechanism of Galactic spiral structure, but still controversial in
observation.
%
%It partially due to the lack of a sample of oldest stars
%with accurate distances.
%
To resolve this issue, high precision distance measurements for a
large number of old stars are necessary.
By taking advantage of the latest dataset of {\it Gaia} Early Data
Release 3~\citep[hereafter {\it Gaia} EDR3,][]{Gaiaedr3+2021}, it is
now possible to identify a large number of red clump stars with
high-precision parallaxes, and depict the nearby P-arms traced by older
stars (ages $\sim$2~Gyr).
In this work, we present the results and compare them with the SF-arms
indicated by gas and young objects, aim to better understand the
properties of spiral structure in the Milky Way.

This work is organized as follows. The sample of red clump stars is
identified and described in Sect.~\ref{sec2}. Their distribution on
the Galactic disc and the properties of P-arms are given in
Sect.~\ref{sec3}. Discussions and Conclusions are in
Sect.~\ref{sec4}.

%--------------------------------------------------------------------
\section{Sample of red clump stars}
\label{sec2}

In the late evolutionary stage of low mass ($<$ 2.2 M$_\odot$) stars,
they will undergo core-helium and shell-hydrogen burning, and evolve
to the phase of red clump stars, which have a narrow range of
intrinsic luminosities and colors~\citep{Girardi_2016}.
Red clump (RC) stars account for a large proportion of the cataloged
\textit{GLIMPSE} stars~\citep{Churchwell+2009}.
As discussed in Sect.~\ref{intro}, the observational evidence for the
two-armed picture of Galactic P-arms is primarily from the statistics
of the surface density of \textit{GLIMPSE} RC stars as a function of
the Galactic longitude~\citep{Churchwell+2009}.
The RC K-giants observed in K-band are always dominated by objects
with ages of around 2~Gyr~\citep{Salaris_2012}.
With a median absolute magnitude $M_\mathrm{K_S}$ =
$-$1.61~mag~\citep{Alves_2000,Cabrera-Lavers+2007} and $M_\mathrm{G}$
= 0.5~mag~\citep{Gaiadr2+2018,Andrae+2018}, RC stars within a few kpc
of the Sun can be detected by {\it Gaia}, even in the presence of
interstellar dust extinction, making them excellent tracers for the
properties of P-arms in the Solar vicinity.

\begin{figure*}[!ht]
\centering
\includegraphics[width=0.48\textwidth]{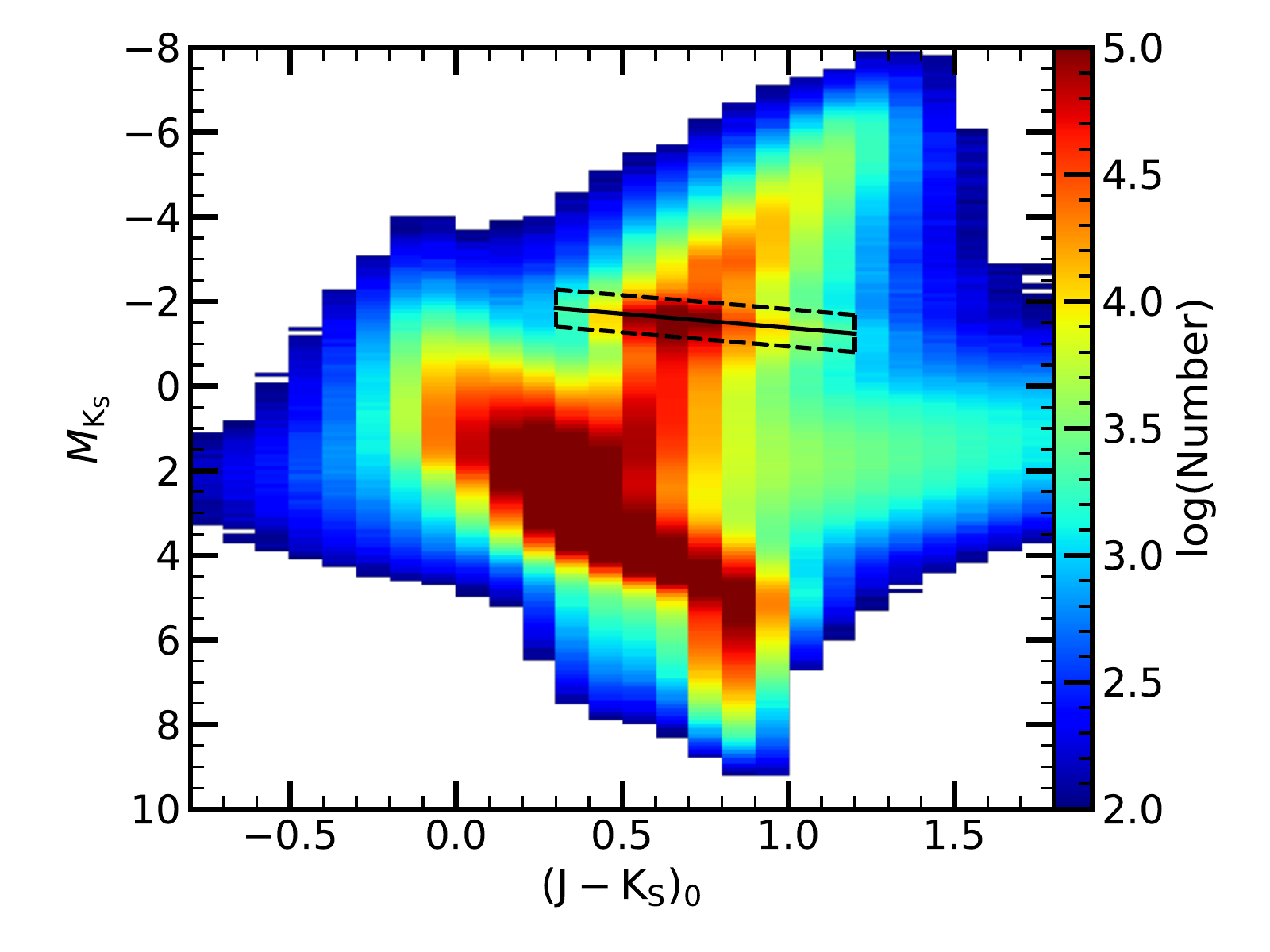}
\includegraphics[width=0.48\textwidth]{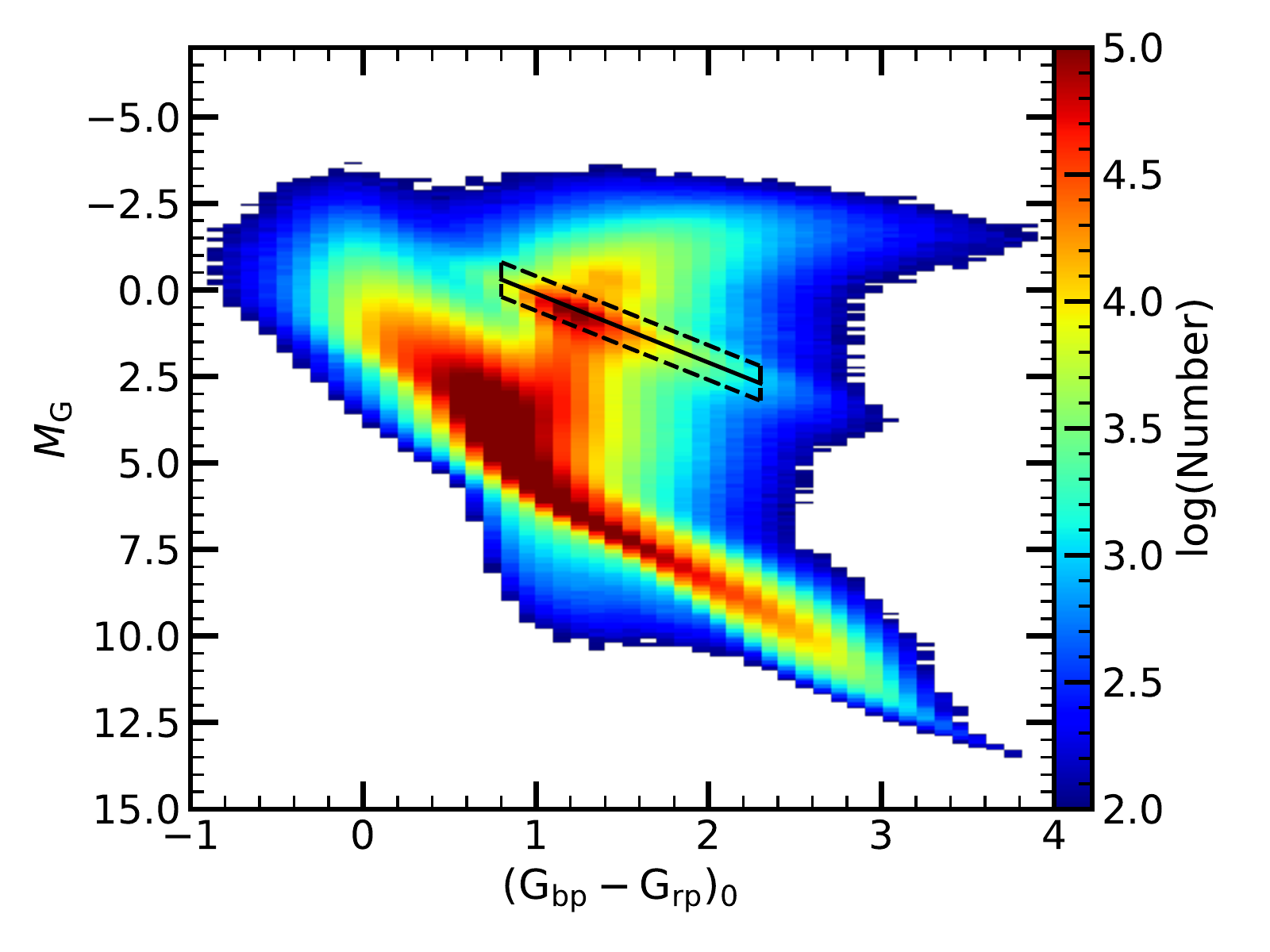}
%\vspace{3.77cm}
\caption{Color-magnitude diagrams: ({\it Left}) $M_{K_S}$ vs. $(J -
  K_S)_0$; ({\it Right}) $M_{G}$ vs. $({G_{bp} - G_{rp}})_0$, for the selected
  {\it Gaia} EDR3$-$2MASS stars with $|Z|$ $<$ 0.3~kpc. To plot the
  diagrams, the cell size is set to be 0.1~mag. Only the cells
  containing more than 100 stars are presented. The black dashed boxes
  indicate the aggregation area of RC stars adopted in this work.}
%  , i.e.,~0.3 $<$ $({J - K_S})_0$ $<$ 1.2,
%  $-$3.725 $<$ 1.5 $\times$ $M_{K_S}$ $-$ $({J - K_S})_0$ $<$
%  $-$2.405, 0.8 $<$ $(G_{bp} - G_{rp})_0$ $<$ 2.0, and $-$2.4 $<$
%  $M_{G}$ $-$ 2 $\times$ $(G_{bp} - G_{rp})_0$ $<$ $-$1.4. }
\label{fig1_HR}
%\figurenum{A. 5}
\end{figure*}

\subsection{Sample selection}

The Two Micron All Sky Survey~\citep[2MASS,][]{Skrutskie+2006}
observed the entire celestial sphere in three NIR bands, i.e. J
(1.25~$\mu$m), H (1.65~$\mu$m), and K$_\mathrm{S}$ (2.16~$\mu$m).
The {\it Gaia} EDR3 has obtained astrometric parameters for about
1.468 billion stars~\citep{Gaia+2016,Gaiaedr3+2021}, with a typical parallax
uncertainty of 0.02$-$0.03~mas for G $<$ 15, and 0.07~mas at G$=$17.
% which enables reveal the local spiral structure in a wide rang with
% $\sim$ 5 kpc.
%
By combining the astrometric/photometric data of {\it Gaia} EDR3 and
the NIR photometric data of 2MASS, a large number of RC stars can be
identified.

As the first step, we extract the sources by using the official {\it Gaia} EDR3$-$2MASS cross-match best neighbor table~\cite[Marrese et al., in prep.,\footnote{\url{https://archives.esac.esa.int/gaia}}][]{Marrese+2019}.
For the 2MASS sources in the Point Source
Catalog~\citep{Skrutskie+2006}, only those stars with $J$ and $K_S$
band magnitudes brighter than 14~mag are retained in consideration of
the sensitivity.
Considering the completeness of the cross-matched sample of {\it Gaia}
EDR3$-$2MASS~\citep{Bennett_Bovy_2019}, the {\it Gaia} sources with G
band magnitude brighter than 17~mag are selected.
It is also required that the sources should have relative parallax
uncertainties $\varpi$/$\sigma_\varpi >$ 5, where $\varpi$ and
$\sigma_\varpi$ are the parallax and its uncertainty without
systematic error, respectively.
Then, the distance $d$ is simply calculated through $d~=~1/(\varpi -
\varpi_0)$, and $\varpi_0$ represents the systematic zero point offset
of the {\it Gaia} EDR3 parallaxes~\citep{Lindegren+2021}.
As we focus on the properties of spiral arms in the Galactic disc, the
stars with $|Z|$ $>$ 0.3~kpc are excluded. Here, $Z$ is the vertical
height.

To derive the color-magnitude diagrams for classifying RC stars, extinction parameters for these sources need to be corrected.
In this work, we adopt the extinction parameters corrected with the
\textit{MWDUST}\footnote{\url{https://github.com/jobovy/mwdust}}~\citep{Bovy+2016}
algorithm,
which combines the three dimensional (3D) dust maps
of~\cite{Marshall+2006},~\cite{Green+2019}, and~\cite{Drimmel+2003}.
This algorithm can determine the extinction value at a given Galactic
longitude, latitude and distance.
Besides, the extinction curve with $R_V$ =
3.1~\citep{Cardelli+1989,ODonnell_1994} is adopted. Together with
the mean wavelengths of 2MASS $J$ and $K_S$
bands~\citep{Cohen+2003}, {\it Gaia} $G$, $G_{bp}$, and $G_{rp}$
bands~\citep{Weiler_2018}, we can derive the extinction coefficient
relations, they are $A_{J}$ = 2.5 $A_{K_S}$, $E({J-K_S})$ = 1.5 $A_{K_S}$, $A_{G}$ = 7.2 $A_{K_S}$, and $E({G_{bp} - G_{rp}})$ = 0.5 $A_{G}$. The calculated relation of $E({G_{bp} - G_{rp}})$ = 0.5 $A_{G}$ is well consistent with that of \cite{Andrae+2018}.

\begin{figure*}[!ht]
\centering
\includegraphics[width=0.48\textwidth]{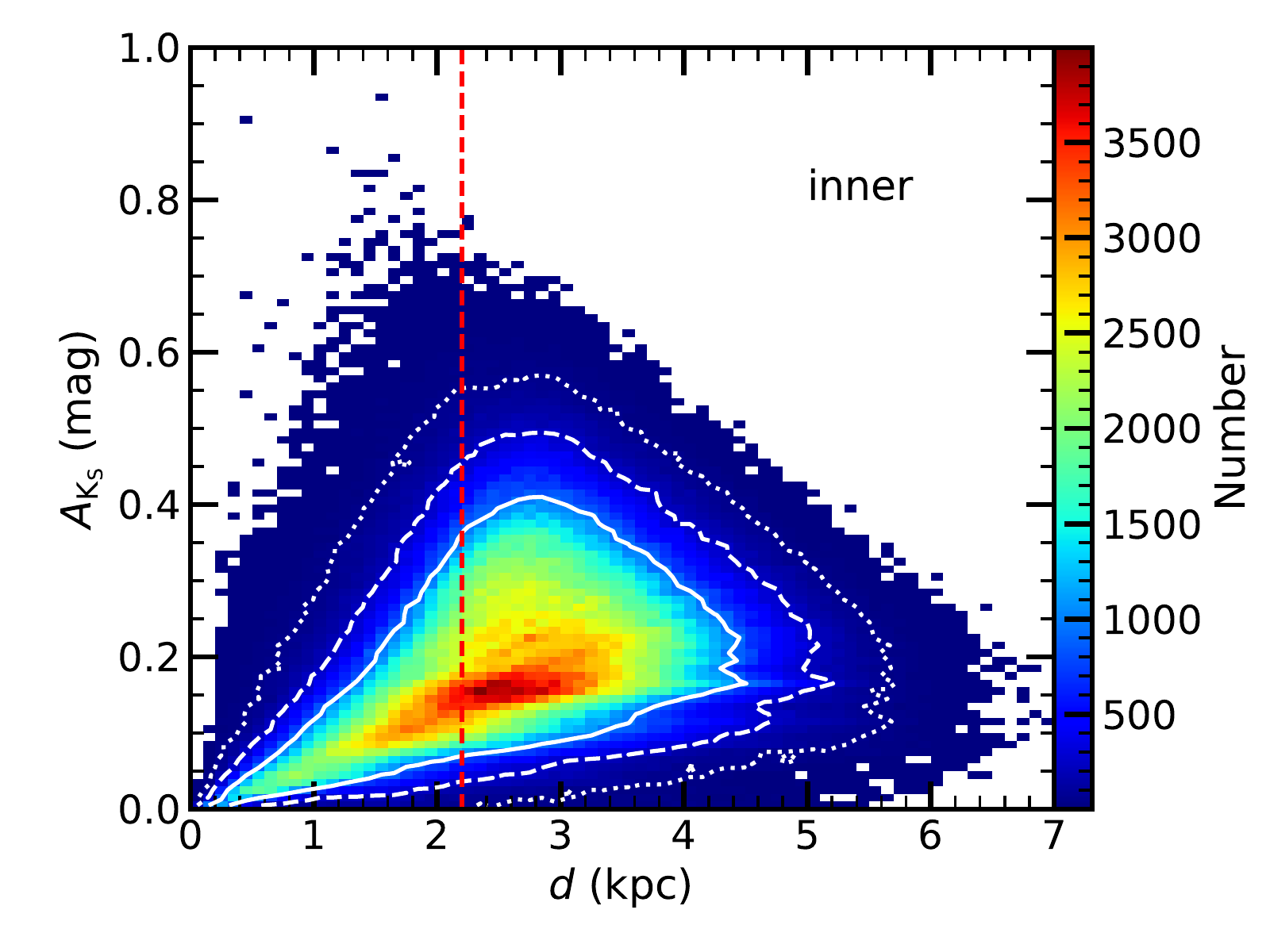}
\includegraphics[width=0.48\textwidth]{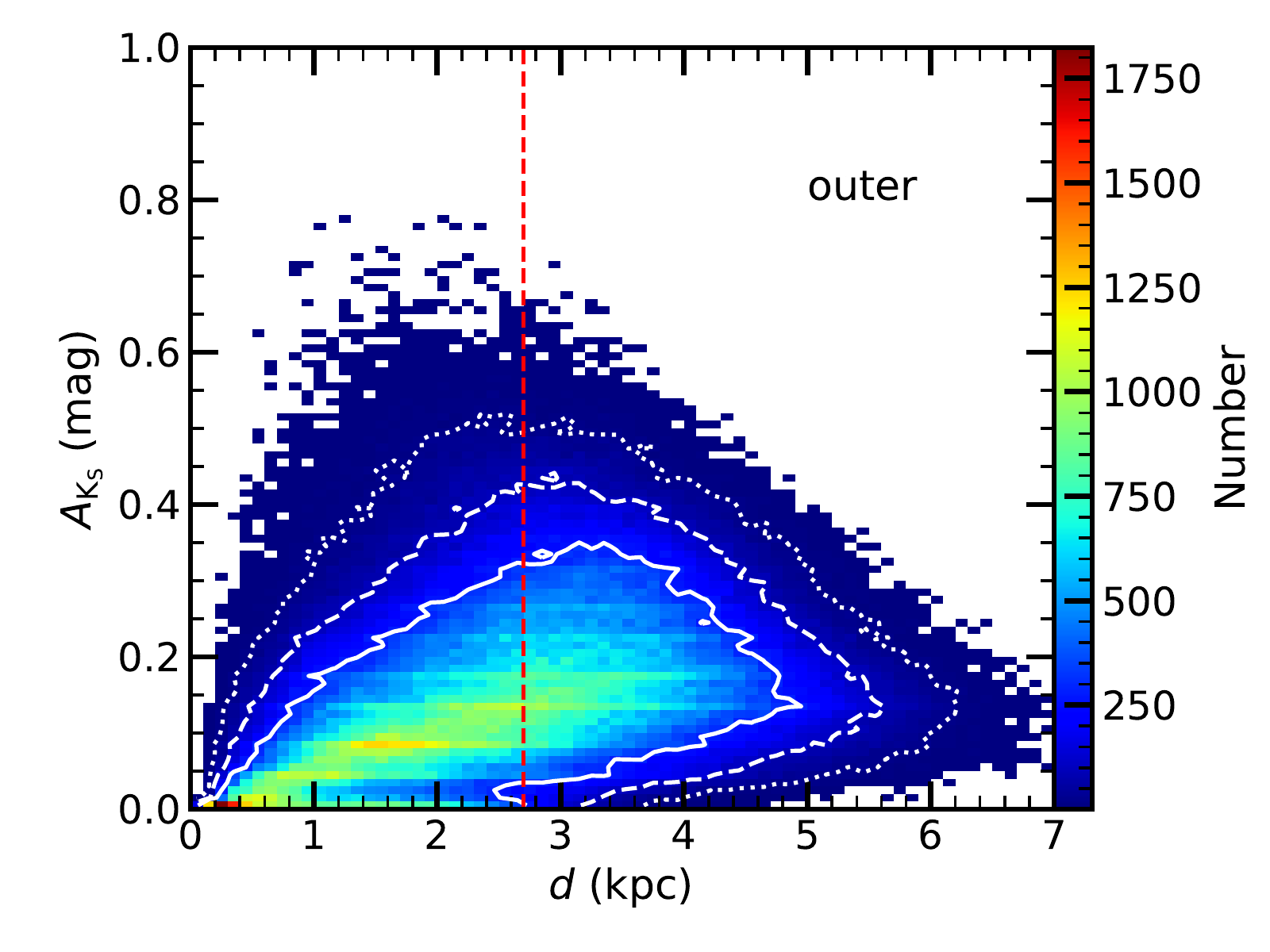}
%\vspace{3.77cm}
\caption{K$_\mathrm{S}$-band extinction of the selected RC stars as a
  function of the distance to the Sun for the inner ({\it Left}) and
  the outer Galaxy regions ({\it Right}). The contours indicated by dotted-, dashed-, and solid-lines enclose the 99\%, 95\%, and 80\% of the RC stars in our sample, respectively.}
\label{fig2_Aks}
%\figurenum{A. 5}
\end{figure*}

The absolute color and magnitude for a matched {\it Gaia} EDR3$-$2MASS
source can be calculated via:
\begin{align}
  & M_{K_S}~=~m_{K_S}~+~5~\times~\log{(\varpi - \varpi_0)}~-~10~-~A_{K_s},\\
  & ({J - K_S})_0~=~({J - K_S})_{obs}~-~E({J-K_S}),\\
  & M_{G}~=~m_{G}~+~5~\times~\log{(\varpi - \varpi_0)}~-~10~-~A_{G}, \\
  & ({G_{bp} - G_{rp}})_0~=~({G_{bp} - G_{rp}})_{obs}~-~E({G_{bp} - G_{rp}}).
\end{align}

The absolute color-magnitude diagrams for the selected sample is
presented in Fig.~\ref{fig1_HR}.
In the diagrams, the RC stars are relatively isolated from other types
of stars, hence can be easily identified~\citep{Girardi_2016}.
It should be mentioned that, the actual extinction of a specific
star depends on its spectral type and also the extinction itself
\citep[e.g.,][]{Jordi+2010}, which may deviate from the statistical
extinction distribution of \textit{MWDUST}. This effect results in
the stretch of aggregation area of RC stars in the color-magnitude diagrams.
% Considering that the extinction of a single star may deviate from
%the statistical extinction distribution of \textit{MWDUST}, the will
%be stretched.
Taking this into account, we select the RC stars by using their expected colour dependency and also the extinction uncertainty. The red clump stars typically have a median absolute magnitude $M_{K_S,RC}$ = $-$1.61 mag, with a Gaussian width of $\sigma_{K_S}$ = 0.22 mag \citep[][]{Alves_2000,Cabrera-Lavers+2007}, and an intrinsic colour of $(J - K_S)_{0,RC}$ = 0.65 mag. Considering a 2$\sigma$ confidence level, the stars with $M_{K_S}$ falling within [$-$2.05,$-$1.17]~mag are selected as candidates of red clump stars. Then, the sample selection box is obtained in consideration of the properties of RC stars along the extinction direction as shown in Fig.~\ref{fig1_HR}, where the black solid line corresponds to the offset of median magnitude due to the extinction uncertainty.
In this work, the RC stars are selected by requiring that 0.3 $<$ $(J
- K_S)_0$ $<$ 1.2, $-$3.725 $<$ 1.5 $\times$ $M_{K_S}$ $-$ $(J -
K_S)_0$ $<$ $-$2.405, 0.8 $<$ $(G_{bp} - G_{rp})_0$ $<$ 2.3, and
$-$2.4 $<$ $M_{G}$ $-$ 2 $\times$ $(G_{bp} - G_{rp})_0$ $<$ $-$1.4, as
indicated by the black dashed boxes in Fig.~\ref{fig1_HR}.
In total, 2,685,075 candidates of RC star are selected by these criteria.
To evaluate the influence of the adopted 3D extinction model, we also reproduce the sample of RC stars with a different 3D extinction algorithm of \cite{Amores+2021}, and found that the derived structure features in the overdensity maps of RC stars are consistent.

%We also reproduce the sample of RC stars with the 3D
%  extinction models of Amores et al. (2021) and Chen et al.(2019),
%
%, we also
%.

To inspect the reliability of the sample, we cross-matched the
selected candidates of RC stars with the catalogue
of~\cite{Sanders_Das_2018}. \cite{Sanders_Das_2018} obtained the
parameters of distances, masses and ages for about 3 million stars
from {\it Gaia} DR2, based on the spectroscopic parameters derived
from several large spectroscopic surveys. We found that there are
92,525 matched sources, 87,623 ($\sim$ 94.7\%) of them are giant stars
with a typical age of $\sim$3~Gyr. It supports that the overwhelming
majority of our selected sources are indeed old stars.

\begin{figure*}[!ht]
\centering
\includegraphics[width=0.32\textwidth]{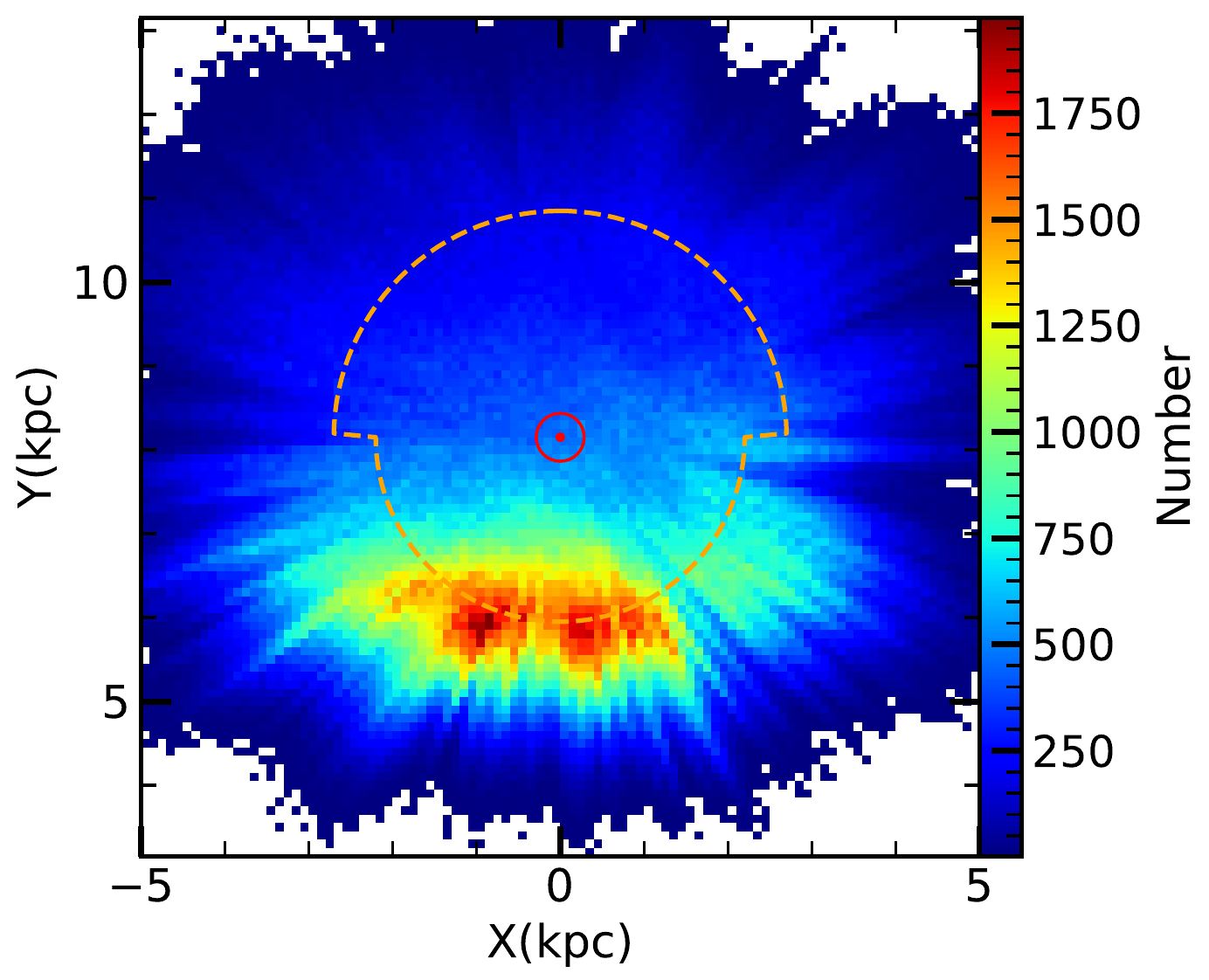}
\includegraphics[width=0.32\textwidth]{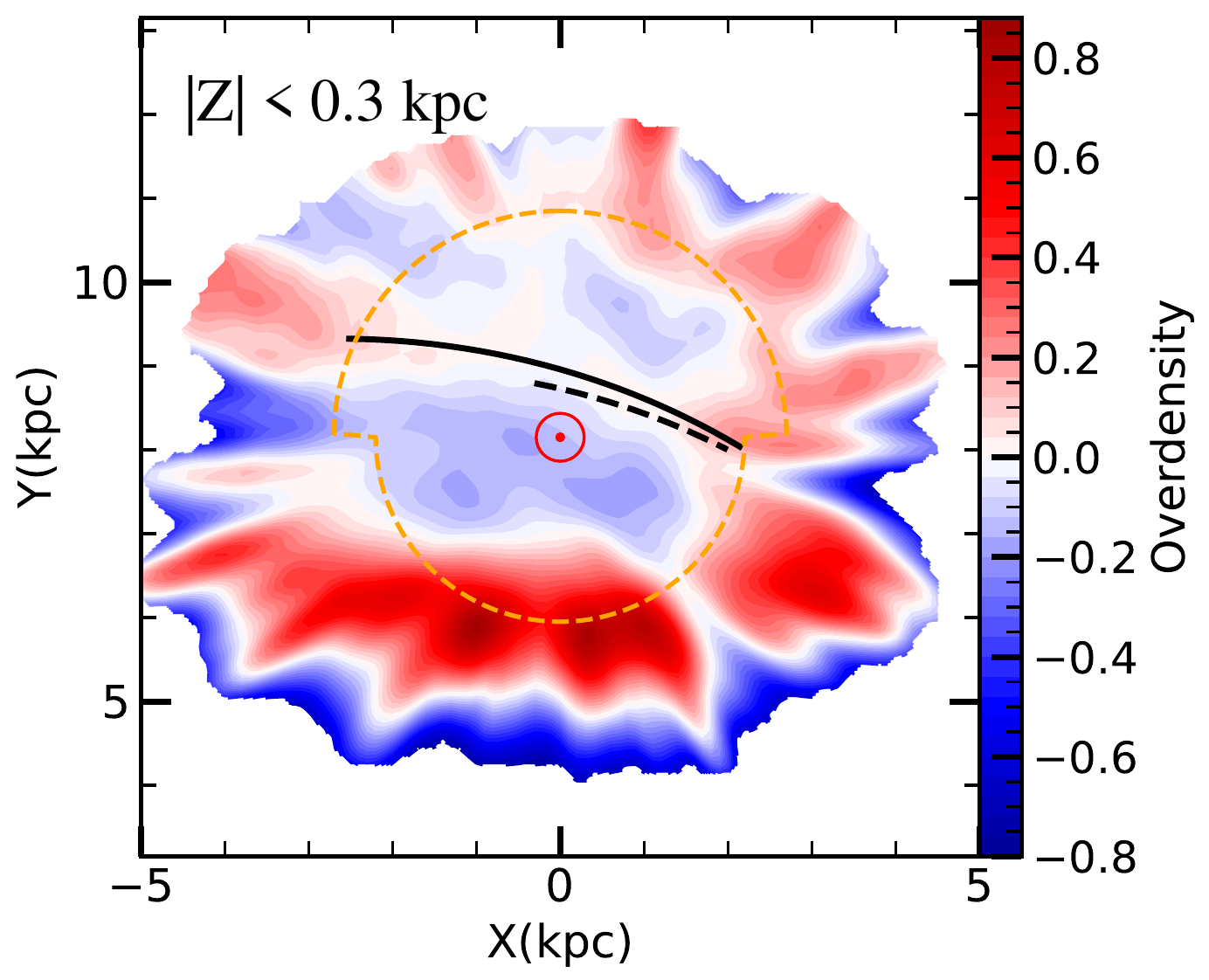}
\includegraphics[width=0.32\textwidth]{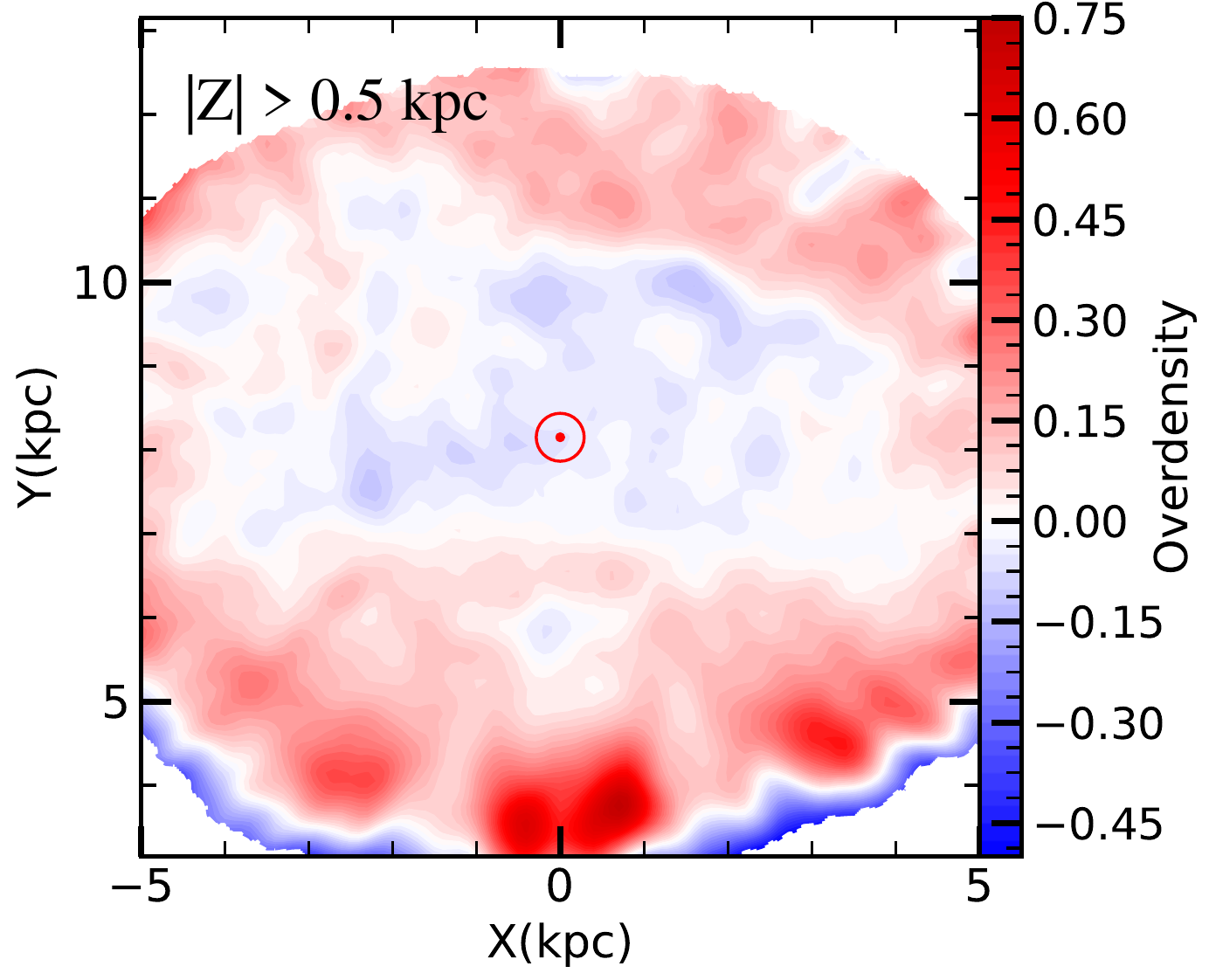}
%\vspace{3.77cm}
\caption{{\it Left}: number density distribution of RC stars
  projected onto the Galactic plane. {\it middle}: overdensity
  distribution of the RC stars with $\varpi$/$\sigma_\varpi$ $>$ 5
  and $|Z|$ $<$ 0.3~kpc, calculated with a local density scale length
  of 0.3~kpc and a mean density scale length of 2~kpc. Only the points
  with $\Sigma(X,Y)$ $>$ 0.003 are presented in consideration of
  statistics. {\it Right}: simlar with {\it middle} panel, while presents the overdensity distribution of the RC stars with $\varpi$/$\sigma_\varpi$ $>$ 5
  and $|Z|$ $>$ 0.5~kpc. The Galactic centre is at (0,0)~kpc, and the Sun (Sun
  symbol) is located at (0, 8.15)~kpc. The black solid line presents the fitted Local P-arm (see detail in Sect.~\ref{sec3_2}) and the dashed line presents the Local Arm fitted by~\cite{Miyachi+2019}. The orange dash line contour the region that RC stars are completed.}
\label{fig3_xy}
\end{figure*}

\subsection{Sample completeness}
\label{com}

An important issue in uncovering the spiral arms with optically or NIR
selected stars is the sample completeness. It depends on the dust
extinction, which is related with the spiral structure to some degree.
In Fig.~\ref{fig2_Aks}, we present the extinction $A_{K_S}$ of the
selected RC stars in this work.
Considering that the dust extinction in the inner Galaxy region
($-90^\circ~<~l~<90^\circ$) is more severe than the outer Galaxy
($90^\circ~<~l~<270^\circ$), we simply divide the RC sample into two
different parts.
For the vast majority of the selected RC stars ($>$ 99\%), the extinction in the direction of the inner Galaxy is less than 0.54
mag, and less than 0.50 mag for the outer Galaxy direction.
For more distant stellar objects, extinction values are bound to be greater than the maximum of the current sample.The completeness of samples drops at the farther distance.
Considering that {\it Gaia} dataset of stars is complete down to at
least $G=$~17~mag, as shown in Fig.~\ref{fig2_Aks}, the RC stars
selected in this work is believed to be complete at
$d\lesssim$~2.2~kpc for the inner Galaxy direction, and at
$d\lesssim$~2.7~kpc for the outer Galaxy regions.
At $G=$ 17 mag, the median error of the {\it Gaia} EDR3 parallax is around 0.07. At 2.2$/$2.7~kpc the parallax
uncertainties (6.5$/$5.3) is large than 5. Fig.~\ref{fig2_Aks} also shows the situation that the sample completeness is insufficient. The
high extinction data is missing beyond the distance far from
the sample completeness (the red dashed lines in Fig.~\ref{fig2_Aks}).
Therefore, at present, we only discuss the distribution of old stars within the limits as indicated by sample completeness limit.
Hence, the dataset of RC stars is expected to depict the properties of
the nearby arm-segments of the Local arm.
At larger distances (e.g., $d\gtrsim3$~kpc), the sample is not
complete, but a large number of RC stars are identified.

%A stellar overdensity method may help us to 
%The Fig.~\ref{fig2_Aks} also shows the situation that the sample
%completeness is insufficient. The high extinction data is missing
%beyond the distance far from the sample completeness (as the red
%dashed line in Fig.~\ref{fig2_Aks}).
%

% With this sample, we can explore the properties of P-arms in the
% vicinity of the Sun.

\section{Distribution in the Galactic plane}
\label{sec3}

The distribution of RC stars projected onto the Galactic plane is
shown in left panel of Fig.~\ref{fig3_xy}.
In the studied area of the Galactic disc, the number density of RC
stars decreases as the increase of Galactocentric distance.
At a Galactocentric distance of $R_\mathrm{GC}\sim$ 6~kpc, where presents an aggregation of RCs, the number density of RCs is about
1500 per 0.01~kpc$^2$, and drops to less than 500 per 0.01 kpc$^2$ at
$R_\mathrm{GC}\sim$ 10~kpc.
%
%It is difficult to exclude the possibility of selection effect based
%on Fig.~\ref{fig3_xy}(a), and connect it to a known spiral arm(s).
%
From the number density distribution, no obvious arm-like features are
discernible.
This is not unexpected, as the features shown in the left panel
of Fig.~\ref{fig3_xy} is completely dominated by the global
density profile of stars in the Galactic disc, which is roughly
exponential.
The P-arms indicated by the overdensities of old stars would be less
significant and concealed beneath the global density profile. In
addition, in comparison to the gas component or young objects, the
velocity dispersion of old stars is larger, the overdensities of old
stars (e.g., indicating the P-arms) causing by any perturbations in
the Galactic disc would be less sharp and more difficult to identify
from observations.
In order to reveal the possible underlying P-arms, a bivariate kernel
density estimator method~\citep{Feigelson_Babu_2012,Poggio+2021} is
adopted. This method may be also helpful to reveal the underlying
overdensity structures outside the completeness limits of the RC
star sample, because the P-arms are expected to leave some marks on
the distribution of the detected distant stars.

Following \citet{Poggio+2021}, the stellar overdensity,
$\Delta_\Sigma$, is calculated by:
\begin{align}
  & \Delta_\Sigma(X,Y) = \frac{\Sigma(X,Y)}{\langle\Sigma(X,Y)\rangle} - 1;\\
  & \Sigma(X,Y) = \frac{1}{N~h^2}\sum^{N}_{i=1}\left[K\left(\frac{X-x_i}{h}\right)~K\left(\frac{Y-y_i}{h}\right)\right];\\
  & K\left(\frac{X-x_i}{h}\right) = \frac{3}{4}\left(1-\left(\frac{X-x_i}{h}\right)^2\right).
\end{align}
Here, $\Sigma(X,Y)$ is the local density at the position ($X, Y$) in
the Galactic disc, $K$ is the kernel function, $h$ is the kernel
bandwidth which is adopted as 0.3~kpc in this work, ($x_i,y_i$)
is the coordinate of the $i$-th RC star, and $N$ is the total number
of RC stars involved in the calculation.
The mean density $\langle\Sigma(X,Y)\rangle$ is calculated in a
similar way as $\Sigma(X,Y)$ but with $h=$~2~kpc. The result is given in the middle panel of Fig.~\ref{fig3_xy}.
As shown in the overdensity maps in middle panel of Fig.~\ref{fig3_xy},
there is a weak arm-like feature extended from $l\sim$~90$^\circ$ to
$l\sim$~243$^\circ$, passed close to the Sun. Most part of this
feature is within the completeness limit of the RC sample
($\sim$~2.2$-$2.7~kpc, see Sect.~\ref{com}), hence is very likely a real structure (as the black solid line shown in the middle panel in Fig.~\ref{fig3_xy}). 
This feature is close to the Local Arm depicted by gas and young objects.
Besides, for comparision, we aslo show the overdensity maps of RC stars away from the Galactic plane ($|Z| > $ 0.5~kpc) in right panel of Fig.~\ref{fig3_xy}. The arm-like structure is only present near the Galactic plane. This may imply that the disk stars show a spiral response under the action of spiral potential for a stellar disc~\citep{Lin_Shu_1964,Kalnajs_1965,Dobbs_Baba_2014}. When the stars are far away from the Galactic plane, the influence of spiral potential weakens, and the overdensity distribution shows a relatively uniform distribution.

%
%
%This arc structure is close to the 
%(see dark cyan solid line in Fig.~\ref{fig3_xy}(b)) of the Milky
%Way, but shows a significant outward drift.
%
%This also implies that there may be a off-set between P-arms and
%SF-arms.
%
In addition, there are obvious overdensity of RC stars in the regions
far away from the Sun.
There may be two different explanations about these overdensity
features. One is the influence of the sample completeness, as many of
these overdensity regions are close to the completeness limit of the
RC stars. 
%
% This possibility could be confirmed or excluded by using a more
% complete sample of RC stars, which is currently not accessible at
% least before the new data release of {\it Gaia}.
%
Another explanation is the spiral structure traced by old stars. 
We look forward to verifying these structures with more complete samples in the future.

\subsection{Comparison with previous works}
\label{sec3_1}

With a sample of turn-off stars of ages $\sim$ 1~Gyr,
\citet{Miyachi+2019} identified a marginal arm-like overdensity in the
Galactic longitude range of 90$^\circ~<~l~<$ 190$^\circ$. This feature
is confirmed with the overdensity map of RC stars of ages $\sim$
2~Gyr. As shown in the middle panel of Fig.~\ref{fig3_xy}, the arm-like
overdensity proposed by~\citet{Miyachi+2019} is part of the intermediate arm-like features identified with RC stars in the same longitude range. 
%, indicating that the  is less dominant.

\begin{figure}[!ht]
\centering
\includegraphics[width=0.49\textwidth]{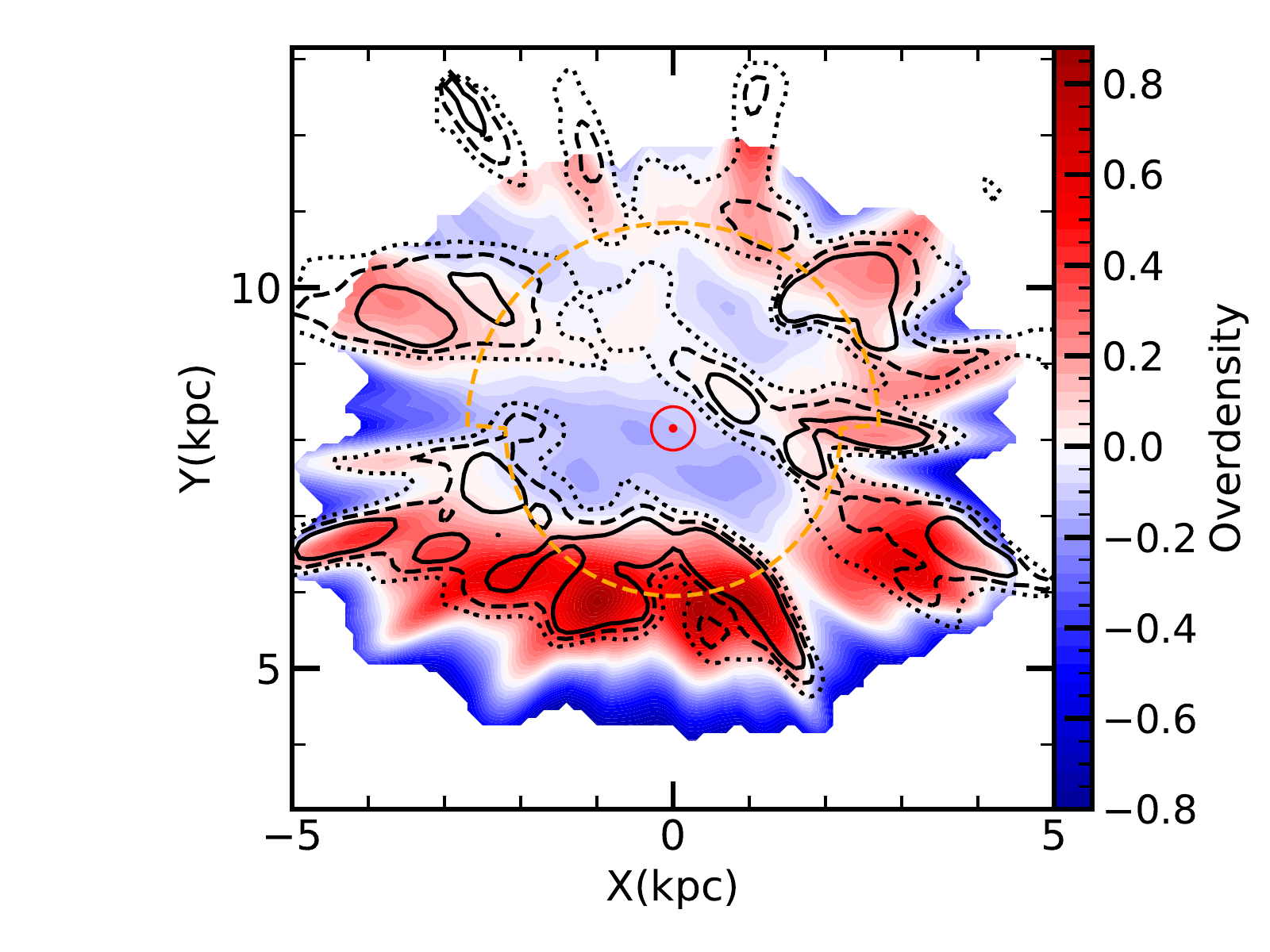}
%\vspace{3.77cm}
\caption{Overdensity distribution of RC stars as the {\it middle} panel
  of Fig.~\ref{fig3_xy}, overplotted with the overdensity map of
  UMS stars~\citep{Poggio+2021}. The black contours represent the
  overdensity levels of 0 (dotted), 0.2 (dashed), and 0.4 (solid) of
  UMS stars, respectively. The orange dash line contour the region that RC stars are completed.}
\label{fig4_poggio}
%\figurenum{A. 5}
\end{figure}

Recently, \citet{Poggio+2021} derived the density distribution of a
large number of upper main sequence stars (UMS stars).
Their upper main sequence stars consist of O-, B- and A-type stars as
discussed in \citet[][]{Poggio+2018}.
We overplotted their results on the overdensity map of RC stars in
Fig.~\ref{fig4_poggio}.
The morphology of the overdensity regions given by \citet{Poggio+2021}
is generally similar to that of RC stars. Three arm-segmemts were proposed by \citet{Poggio+2021} based on the distribution of UMS stars, which is similar to our interpretation about the overdensity map of RC stars.
%also shows that the selection effect and
%completeness of samples are not dominant, but the structure of
%spiral arm itself is dominant.}
%

With UMS stars, \citet{Poggio+2021} identified the Sagittarius-Carina
Arm and also the Scutum-Centaurus Arm (also see their Fig. 2).
%, among them a segment of the Sagittarius-Carina Arm is located in the complete region of RC stars. There is also presents an over-density as shown in the overdensity map of RC stars, while an more over-density is shown outside the complete region.
RC stars also presents over-density in the Scutum-Centaurus Arm and Perseus Arm identified by~\citet{Poggio+2021}, while it is outside the complete region. We only show the coincidence of this structure here.

Due to the limitation of sample completeness at present, we only compare RC stars and UMS stars in area of high completeness (orange contour region in Fig.~\ref{fig4_poggio}). 
In the overdensity map of UMS stars, the arm-like feature near the Sun is more obvious.
We speculate that, the main cause is that the sample used by
\citet{Poggio+2021} is a combination of stars covering a wide range of
ages, e.g., the typical lifetime for a B-type star is $\sim$ 0.1~Gyr, and
$\sim$ 1~Gyr for a A-type star, comparable to that of RC stars
($\sim$ 2~Gyr). As discussed in \citet[][]{Poggio+2018}, approximately
55\% of their UMS stars are OB type, and 40\% are A-type stars.
Therefore, the overdensity map of UMS stars seems to present combined features of young objects and old stars. 
The older OCs (e.g., ages $>$ 0.2 Gyr) show a different distribution than the populous OCs with ages of tens of million years~\citep{Hao+2021}, this may indicate the dynamic evolution or state of star move out of spiral SF-arms.
For the purpose of revealing the possible differences between
SF-arms and P-arms in the Milky Way, the RC stars adopted in this work
would be a suitable dataset for illustrating the properties of
P-arms.

\begin{figure}[!ht]
\centering
\includegraphics[width=0.49\textwidth]{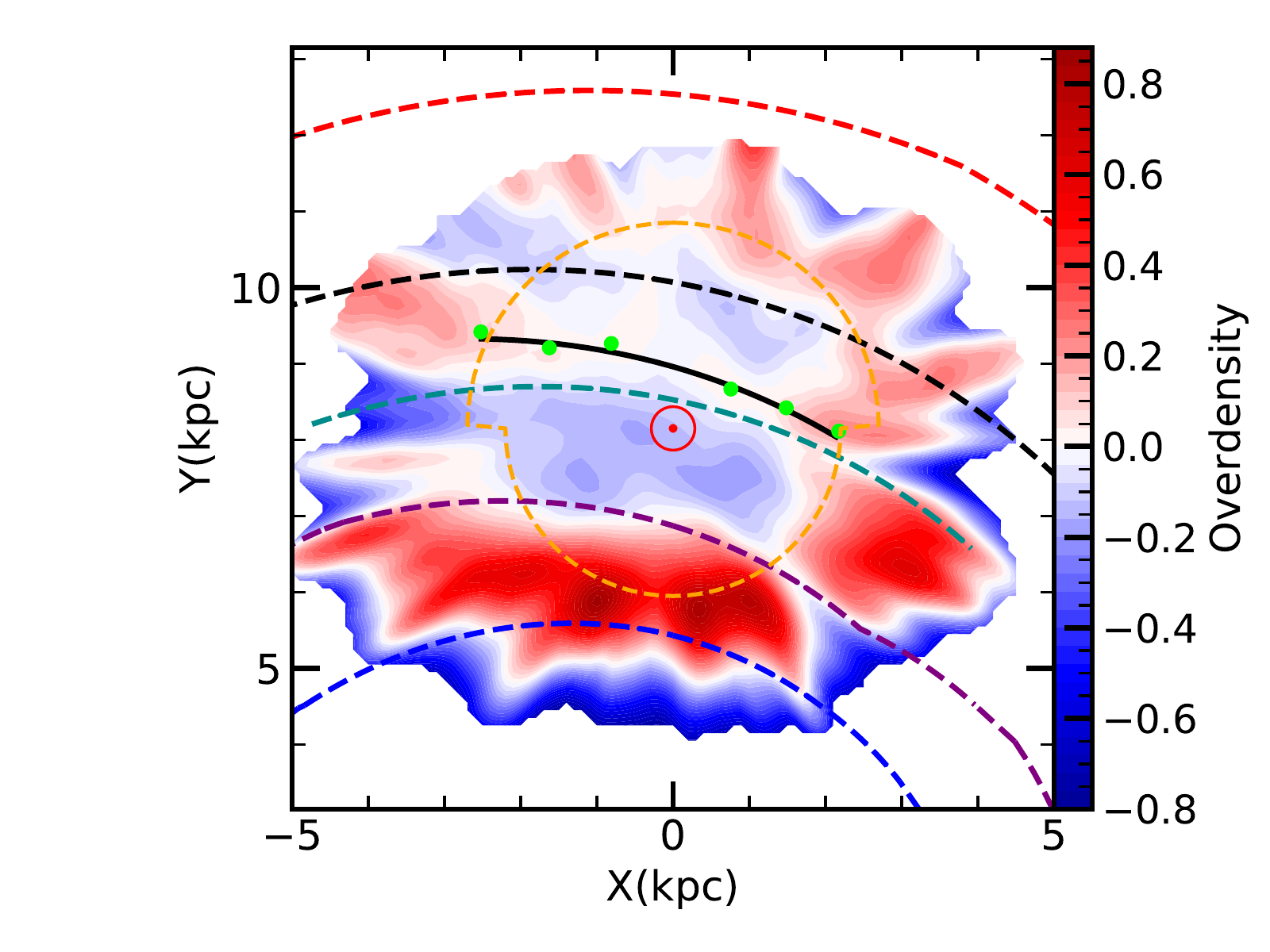}
%\vspace{3.77cm}
\caption{Same as the {\it middle} panel of Fig.~\ref{fig3_xy}, overplotted with the fitted Local P-arm (black solid line) and the dashed lines present the SF-arm fitted by HMSFR masers~\citep{Reid+2019}. The SF-arm of the Outer Arm (red), the Perseus Arm (black), the Local Arm (dark cyan), the Sagittarius-Carina Arm (magenta), and the Scutum-Centaurus Arm (blue) are indicated with different colors. The green dots present the local maxima of the Local P-arm in different $\beta$. The orange dash line contour the region that RC stars are completed.}
\label{fig5_local}
%\figurenum{A. 5}
\end{figure}

\begin{figure*}[!ht]
\begin{minipage}[b]{0.27\textwidth}
\includegraphics[width=1\textwidth]{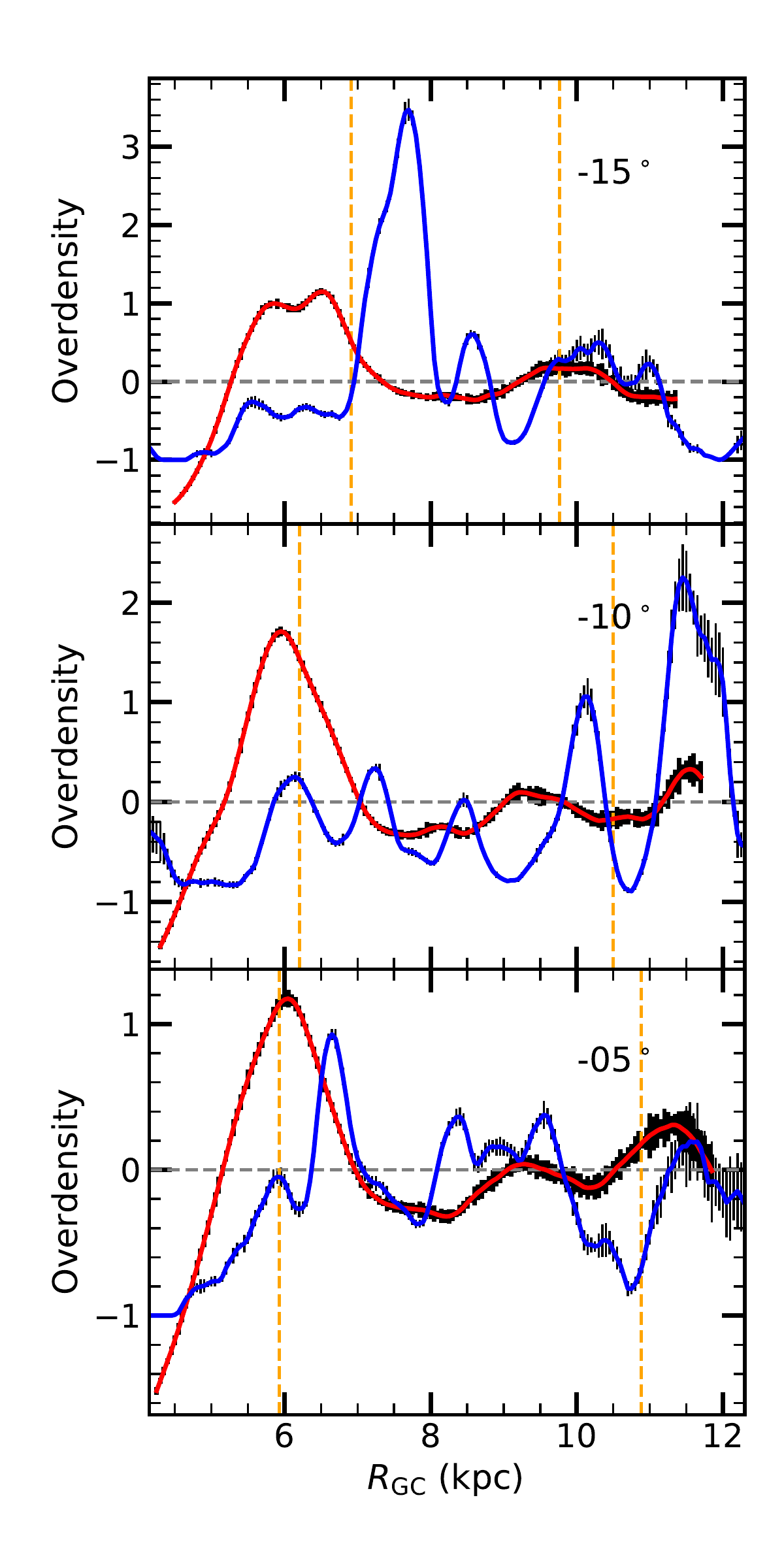}
\end{minipage}
\begin{minipage}[b]{0.42\textwidth}
\subfigure[]{\includegraphics[width=1\textwidth]{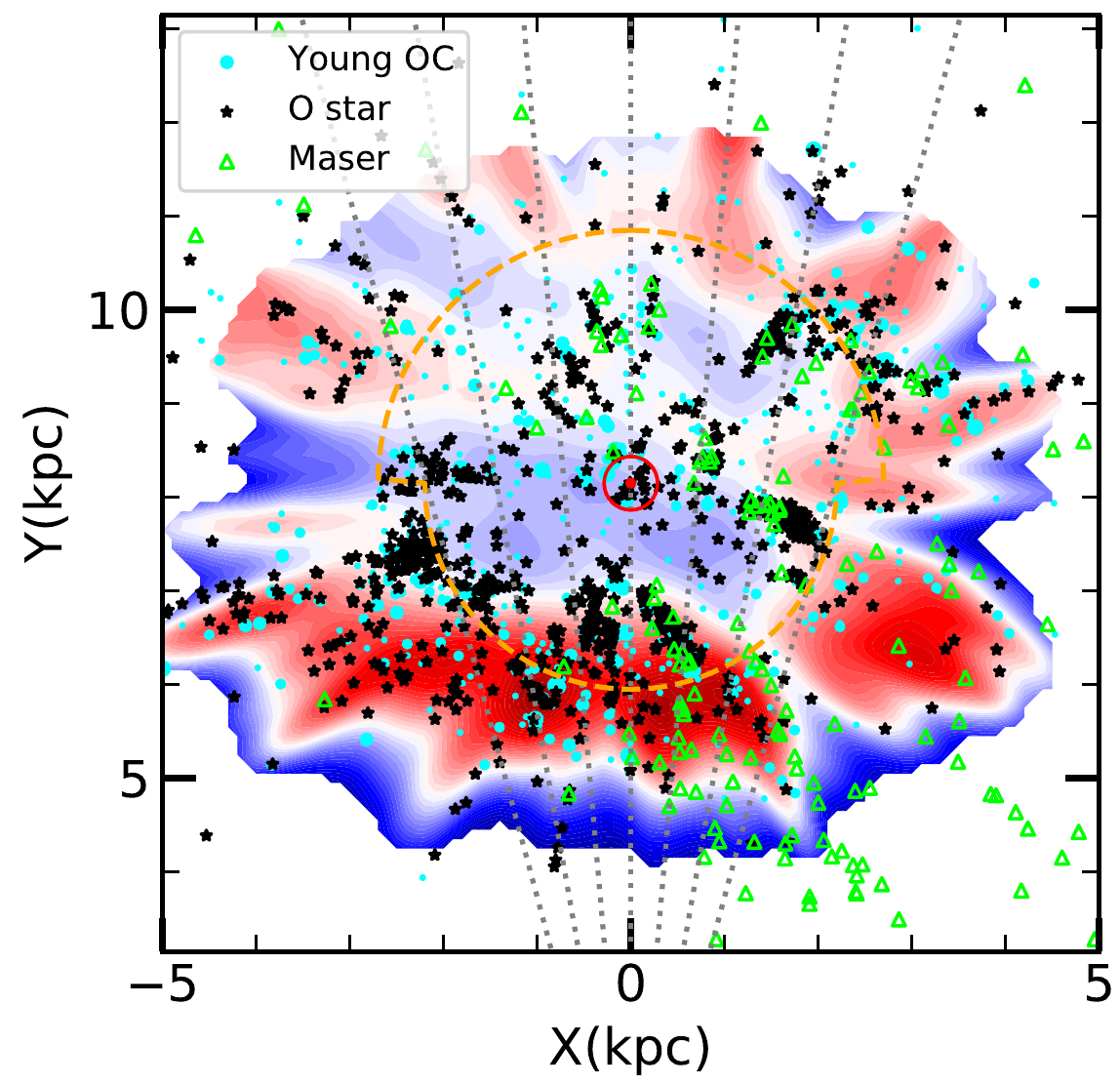}}
\centering
\includegraphics[width=0.61\textwidth]{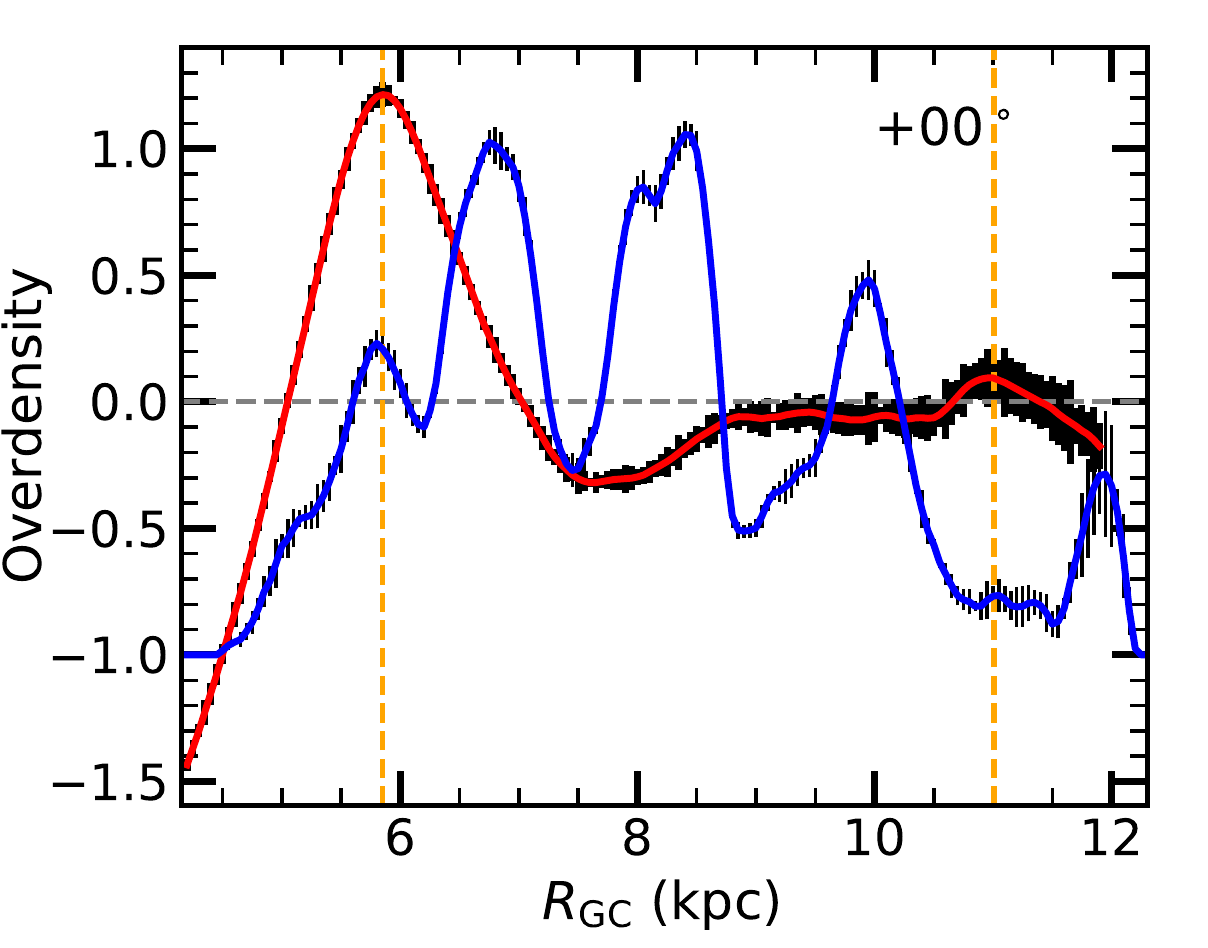}
\end{minipage}
\begin{minipage}[b]{0.27\textwidth}
\includegraphics[width=1\textwidth]{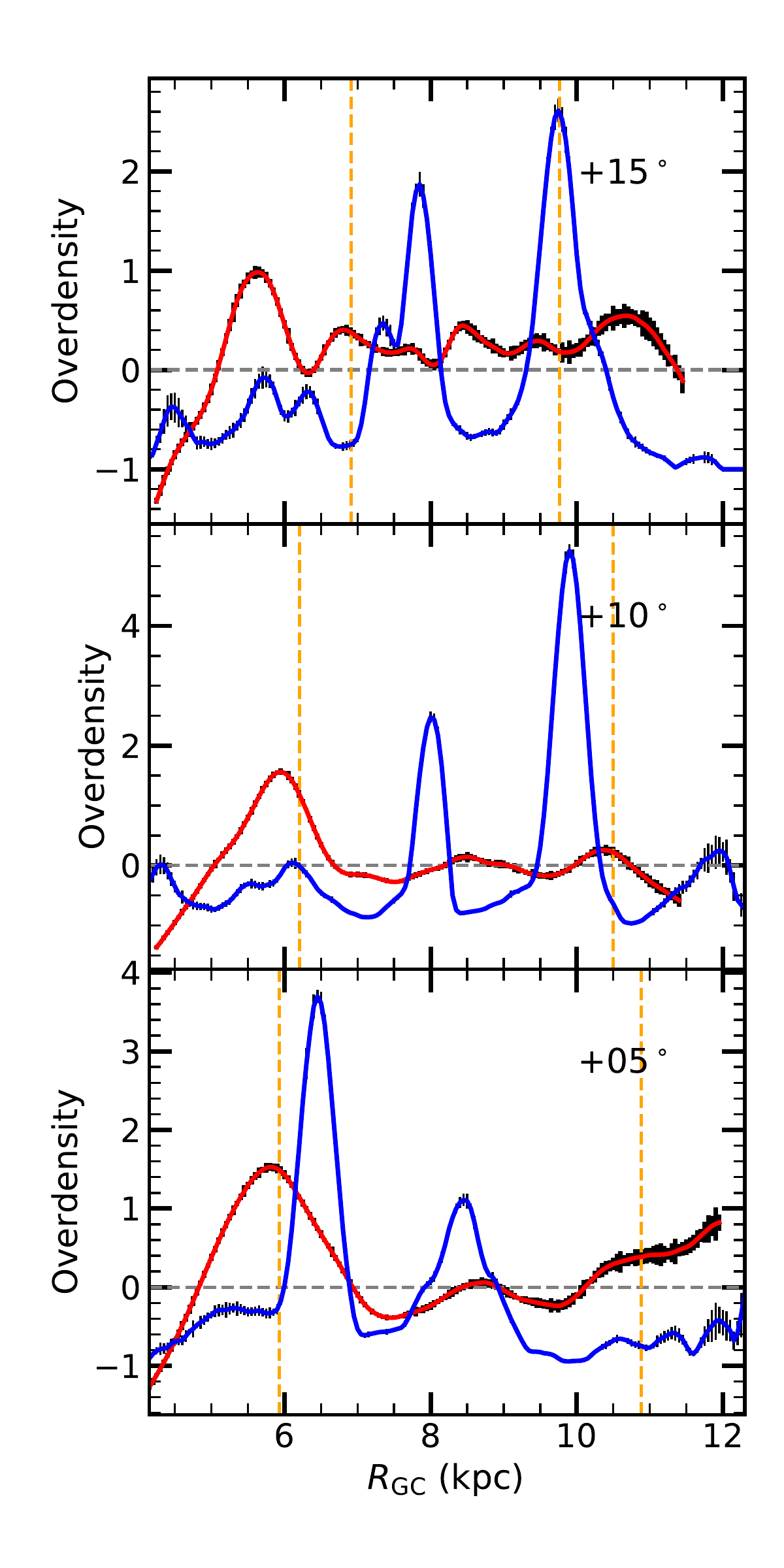}
\end{minipage}
\caption{{\it Panel a}: The overdensity map of RC stars, but overlaid with the distributions of young OCs, massive O-type stars and HMSFR masers having accurate parallax distances. Profiles of the overdensity of RC stars as a function of the Galactocentric distance ($R_\mathrm{GC}$) along different slices ($-15^\circ$, $-10^\circ$, $-5^\circ$, $0^\circ$, $+5^\circ$, $+10^\circ$, and $+15^\circ$) are presented. Each slice corresponds to a constant Galactocentric azimuth angle $\beta$. Error bars show the uncertainties calculated by 100 times of resamples. Additionally, a scaled overdensity profile of young objects as a function of the Galactocentric distance (blue solid line) is also shown for each $\beta$ slice, in order to make a comparison with that of the RC stars (red solid line, the overdensity values and errors of RC stars in this figure are multiplied by a factor of 2). Between the two orange  vertical dashed lines is the complete interval of the sample of RC stars.}
\label{fig6_slice}
%\figurenum{A. 5}
\end{figure*}

\subsection{Local P-arm revealed by RC stars}
\label{sec3_2}

A discernible feature passes close to the Sun, with a over-density in the azimuth angle range of $-15^\circ \lesssim \beta \lesssim15^\circ$. 
The overdensity regions from (X,Y)$\sim$(2,8)~kpc to ($-$2,9)~kpc are close to the Local Arm, and also possibly related with an
arm-segment.
As this feature is also shown by \citet{Miyachi+2019} and \citet{Poggio+2021} with different stellar samples, it is very likely a real arm-like structure.
This arm-like feature is probably corresponding to the segments of the Local P-arm traced by older stars of $\sim$2~Gyr.
The Galactocentric distance of local maxima corresponding to the Local P-arm along different azimuth angle (i.e., $\beta$ = $-15^\circ$, $-10^\circ$, $-5^\circ$, $+5^\circ$, $+10^\circ$, and $+15^\circ$) are 9.75, 9.35, 9.3, 8.7, 8.55, 8.40~kpc, respectively. While at $\beta$ = $0^\circ$, there seems a spur to connect the Local P-arm and outer over-density arc-like feature. Thus, we discarded the local maxima in this azimuth angle.
Under the assumption that they are the segments of nearby P-arms, we
can use logarithmic spiral-arm model, $\ln(R/R_{ref}) = -(\beta - \beta_{ref})\tan{\psi}$, to fit this structures (as green dots scattered in Fig.~\ref{fig5_local}). The fitted pitch
angle 15$\fdg$9 $\pm$ 1$\fdg$1 for the Local P-arm.
This fitted pitch angle is significantly larger than that of the Local ($\psi \sim 11^\circ$) \citep[][]{Reid+2019}, as shown in Fig.~\ref{fig5_local}.
By researching turn-off stars in the Solar vicinity,~\cite{Miyachi+2019} found that the pitch angle of the P-arm is slightly larger than the SF-arm. The results for RC stars also consist with this result. The Milky Way's old stars may exhibit a more loose spiral structure.

\subsection{Comparison with SF-arms}
\label{sec3_3}

As discussed above, the two different components of spiral arms
(SF-arms and P-arms) in a galaxy may be decoupled in their positions.
In order to reveal the possible differences between the identified
P-arms and the SF-arms of the Milky Way, we compare the distribution
of young objects with the overdensity map of RC stars.

The segments of SF-arms traced by young objects (i.e. with ages less than $\sim$ 20~Myr) have been updated and well studied in the past few years, by using HMSFR masers, OB-type stars and young open clusters
with accurate parallax distances.
For these young objects, we adopted the data given by
\citet[][]{Reid+2019}, \citet{Xu+2021} and \citet{Hao+2021}.  The samples of young objects are also suffered from the completeness issues. The available dataset of young objects have been used to trace the SF-arms in the Galaxy disc as far as about 5~kpc from the Sun.
The results are shown in Fig.~\ref{fig6_slice}.
The prominent feature is that the segments of P-arms traced by RC
stars (ages $\sim$ 2~Gyr) are obviously deviated from that of SF-arms, especially for the intermediate arm-like feature, which is not suffered from the sample completeness issue significantly.
To indicate it clearly, a bivariate kernel density estimator method is
also applied to the young objects, then used to calculate the
overdensity profile of young objects as a function of the
Galactocentric distance along the slices with azimuth angle of $-15^\circ$, $-10^\circ$, $-5^\circ$, $0^\circ$,
$+5^\circ$, $+10^\circ$, and $+15^\circ$.
Take the result of $\beta$ = $+10^\circ$ as an example (see
Fig.~\ref{fig6_slice}).

The young objects gather in the areas with $R_\mathrm{GC}$ = 6.2, 8.0 and 9.9~kpc, correspond to the SF-arms of Sagittarius-Carina, Local and Perseus, respectively. 
In comparison, the overdensity of RC stars appear near $R_\mathrm{GC}$ = 8.5~kpc within the area of high
completeness. As mentioned in Sect.~\ref{sec3_2}, this area may correspond to the Local P-arm. There is an obvious outward drift between the Local P-arm and SF-arm. In previous works, the knowledge of P-arms of the Milky Way is primarily from analyzing the profiles of integrated NIR or FIR emission/sources as a function of Galactic longitude
\citep[][]{Drimmel_2000,Drimmel_Spergel_2001,Benjamin+2005,Churchwell+2009}.
The P-arms in the Galactic disc indicate a two-major armed spiral pattern, i.e. the Scutum-Centaurus Arm and the Perseus Arm~\citep{Drimmel_2000,Drimmel_Spergel_2001}.
However, there is no obvious over-density of RC stars in the area of $R_\mathrm{GC}$ = 9.9~kpc, which is correspond to the SF-arms of Perseus Arm. There may be the similar outward shift of the Perseus P-arm. We look forward to a larger and more complete sample to obtain the distribution of Perseus P-arm.
Similar properties are also noticed for the other $\beta$ slices,
implying that the offsets between SF-arms and P-arms in the vicinity
of the Sun are general phenomena.

As shown by studies on some external galaxies \citep[e.g.,][]{Yu_Ho_2018},
the possible offsets between SF-arms and P-arms are generally small,
especially for the regions close to the galaxy co-rotation radius.
It is believed that the co-rotation radius of the Milky Way is close
to the Solar circle~\citep[e.g.,][]{Dias+2019}.
Under the circumstances, the P-arm of the Local arm traced by RC
stars have larger pitch angles than SF-arms, and tend to shift outward
toward the anti-Galactic centre direction. The offset between SF-arm
and P-arm are increased as the arm segments spiral outward.

%

%

% {Especially in the 20\% accuracy sample, the three arc-like features
% are more obvious.  Considering that the results of the samples
% containing more distant RC stars are consistent with the results of
% the present samples, it may be feasible to discuss the structures
% outside the complete region appropriately.  Therefore, in this
% section, a more comprehensive discussion will be made on the current
% overdensity distribution map of RC stars without considering the
% complete region.  }

\section{Discussions and Conclusions}
\label{sec4}

In this work, a large number of RC stars with ages $\sim$2~Gyr is
identified by using the survey data of 2MASS and {\it Gaia} EDR3. In
the Solar vicinity, we found that there are obvious overdensity regions of RC stars. A picture composed of three features could be used to interpret these overdensity features.
Two of the three structures are located outside the completeness limit of the data sample. And one of the three structures probably correspond to the segments of P-arms in the vicinity of the Sun, which have not been well traced in previous works.
This arm-like features possibly correspond to the Local P-arm. 

In comparison to the P-arms, the properties of SF-arms in the Milky
Way have been better illustrated.
By comparing the overdensity distributions of RC stars with that of
young objects, we confirm that there are obvious offsets between the
segments of P-arms and SF-arms in the vicinity of the Sun even though the completeness issue of the sample is considered to some degree. The segment of the Local Arm traced by RC
stars has larger pitch angles than that of SF-arms, tend to spiral
outward toward the anti-Galactic centre direction.
From the studies on external spiral galaxies, it is found that the
SF-arms of galaxies statistically have smaller pitch angles than that
of P-arms \citep[e.g.,][]{Yu_Ho_2018}, hence with a more tightly wound
pattern.
However, the relation between P-arm and SF-arm for the Local seems contradict to the measured arm tangential directions and
corotation radius of the Milky Way.
Based on the density wave theory, the azimuthal offsets between P-arms
and SF-arms will be opposite for the Galaxy regions inside and outside
the corotation
radius~\citep{Roberts_1969,Yuan_Grosbol_1981,Martinez-Garcia+2009}.
The corotation radius for the Milky Way is around the Solar circle.
In the inner Galaxy, the P-arms traced by old stars have also been
found to be outside the corresponding SF-arms by analyzing the arm
tangential directions \citep[e.g.,][]{Hou_Han_2015,Vallee_2014,Vallee_2016,Vallee_2018}, which is similar
to the properties for the Local P-arm noticed in this work.
However, it is not clear whether the systematic
spatial offsets or age pattern exist or not~\citep{
Vallee_2018,He+2021} beyond the outer Galaxy. More tests based on observations are needed and such contradictions are expected to be solved by more complete samples of RC stars and young objects with accurately measured distances.

\acknowledgments
We
would like to thank the anonymous referee for the helpful comments and
suggestions that helped to improve the paper.
This work was funded by the NSFC Grands 11933011, 11873019 and 11988101, the Natural
Science Foundation of Jiangsu Province (Grants No. BK20210999) and the
Key Laboratory for Radio Astronomy.  L.G.H. thanks the support from
the Youth Innovation Promotion Association CAS. 
L.Y.J.
thanks the support of the Entrepreneurship and Innovation
Program of Jiangsu Province.
This work has made use
of data from the European Space Agency(ESA) mission Gaia
(\url{https://www.cosmos.esa.int/gaia}), processed by the Gaia Data
Processing and Analysis Consortium (DPAC, \url{https://www.
  cosmos.esa.int/web/gaia/dpac/consortium}). Funding for the DPAC has
been provided by national institutions, in particular the institutions
participating in the Gaia Multilateral Agreement. This publication
makes use of data products from the Two Micron All Sky Survey, which
is a joint project of the University of Massachusetts and the Infrared
Processing and Analysis Center/California Institute of Technology,
funded by the National Aeronautics and Space Administration and the
National Science Foundation.
%\end{acknowledgements}

\end{document}